\documentclass[prx,aps,twocolumn,showpacs,superscriptaddress,nofootinbib]{revtex4-2}
\usepackage[colorlinks=true,citecolor=blue,linkcolor=magenta]{hyperref}
\usepackage{graphicx}
\usepackage{dcolumn}
\usepackage{bm}
\usepackage{tikz}
\usepackage[bbgreekl]{mathbbol}
\usepackage{amsmath,dsfont}
\usepackage[normalem]{ulem}
\usepackage{color}
\usepackage{booktabs}

\newcommand{\Mmat}{\mathbf{M}}

\newcommand{\stil}{\tilde{s}}
\newcommand{\Atil}{\tilde{A}}
\newcommand{\ttil}{\tilde{t}}

\newcommand{\pvec}{\mathbf{p}}
\newcommand{\kcirc}{\tikz\draw[black,fill=black] (0,0) circle (.5ex);}
\newcommand{\wcirc}{\tikz\draw[black,fill=white] (0,0) circle (.5ex);}

\begin{document}
\title{Universal catastrophe time distributions of dynamically unstable polymers}

\author{Paul B. Dieterle}
\email{dieterle@g.harvard.edu}
\address{Department of Physics, Harvard University, Cambridge, MA 02138, USA}

\author{Jenny Zheng}
\address{Department of Molecular and Cellular Biology, Harvard University, Cambridge, MA 02138, USA}

\author{Ethan Garner}
\address{Department of Molecular and Cellular Biology, Harvard University, Cambridge, MA 02138, USA}

\author{Ariel Amir}
\address{John A. Paulson School of Engineering and Applied Sciences, Harvard University, Cambridge, MA 02138, USA}

\begin{abstract}
Dynamic instability — the growth, catastrophe, and shrinkage of quasi-one-dimensional filaments — has been observed in multiple biopolymers. Scientists have long understood the catastrophic cessation of growth and subsequent depolymerization as arising from the interplay of hydrolysis and polymerization at the tip of the polymer. Here, we show that for a broad class of catastrophe models, the expected catastrophe time distribution is exponential. We show that the distribution shape is insensitive to noise, but that depletion of monomers from a finite pool can dramatically change the distribution shape by reducing the polymerization rate. We derive a form for this finite-pool catastrophe time distribution and show that finite-pool effects can be important even when the depletion of monomers does not greatly alter the polymerization rate.
\end{abstract}

\maketitle

Phenomena differing in their microscopic details can nonetheless share common features. The discovery and description of such universal features is one of the central aims of statistical physics. The identification of universal behaviors is especially valuable in the modeling of living systems, where many microscopic details are unknown or difficult to measure. Amazingly, despite the variety and complexity of biology, some phenomena are observed in many organisms or multiple two domains of life.

One example is the growth of dynamically unstable polymers, in which a quasi-one-dimensional polymer elongates, suffers a catastrophic cessation of growth, depolymerizes, and grows anew. This behavior has been observed in eukaryotic tubulin and a prokaryotic actin homolog \cite{garner2004,mitchison1984,michie2006,desai1997,erb2014}.

The pioneering experiments of Mitchison and Kirschner were the first to demonstrate that the interplay of hydrolysis and polymerization of monomers at the tip of a growing polymer drives it toward catastrophe \cite{mitchison1984}. This observation has been confirmed in several other experiments, also across two domains of life \cite{garner2004,odde1995,padinhateeri2012,michie2006,valiron2001,desai1997,mandelkow1991}.

Despite much experimental and theoretical work, precise mathematical modeling of dynamic instability has proven challenging, owing to the difficulty of probing the microscopic scales of a growing polymer \cite{Michaels2020, Brouhard2015}. It is thus natural to ask which aspects of these models are insensitive to the details. Here, for a broad class of models, we show that the distributions of times to catastrophe (henceforth called the ``catastrophe time distribution") are universally exponential. 

The shape of the distribution is insensitive to noise in polymerization and hydrolysis rates; it is, however, impacted by the depletion of monomers from a finite pool (which has the effect of lowering the polymerization rate). We show that this effect can drastically perturb the exponential catastrophe time distribution, even if it does not greatly alter the polymerization rate. For several models of dynamic instability, depletion of a finite monomer pool results in a catastrophe time distribution with a Gumbel-like form.

\section*{Model construction}

To model the stochastic growth and hydrolysis of a growing one-dimensional polymer, we first construct constant-rate models of each process. In such models, monomers add to the polymer at the tip at a rate, $a$; once added to the polymer, they hydrolyze at a rate, $b$ (see Figs. 1A and 2A). These assumptions appear to be experimentally relevant, owing to the fact that dynamically unstable polymers elongate and hydrolyze at roughly constant rates \cite{fygenson1994,bowne2013,brun2009,padinhateeri2012,garner2004,odde1995,janson2003,walker1988,hill1984,zong2006}.

To deduce the catastrophe time distribution for such models, we must first specify the states of a polymer that result in catastrophic depolymerization. We will begin by grounding our analysis in two specific models, then deduce the universal aspects of all such constant-rate models.

\begin{figure}[t!]
    \centering
    \includegraphics[width=8.6cm]{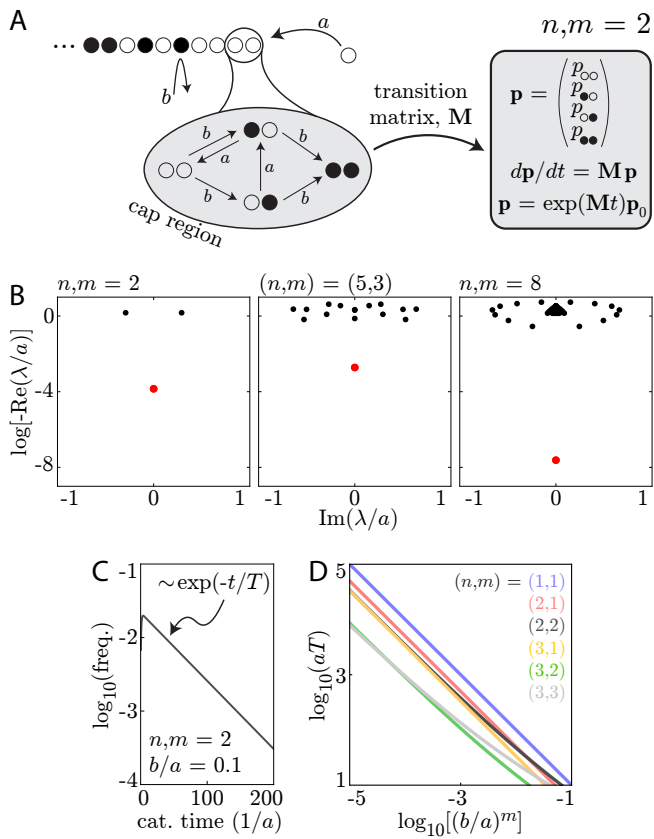}
    \caption{\textbf{Cap model dynamics.} \textbf{A:} Schematic of the cap model. Polymerization and hydrolysis proceed at constant rates -- $a$ and $b$, respectively. The polymer has a cap region of size $n$. When $m$ of the cap region monomers hydrolyze, the polymer catastrophically depolymerizes. Cap models can be described by a Markov Process with transition matrix $\Mmat$. \textbf{B:} Eigenspectrum of the cap model transition matrix, $\Mmat$, shown here for various cap sizes and $b/a=0.1$. In all cases, the eigenspectrum shows a quiescent eigenvalue with small negative real component (red dot). This small negative real eigenvalue dominates the catastrophe dynamics. There is an eigenvalue at zero which is not shown here. \textbf{C:} Catastrophe time distribution for the $n,m=2$ cap model with $b/a=0.1$. A very short transient (which depends on the initial conditions) is followed by an exponential distribution described by a characteristic timescale, $T$. \textbf{D:} Characteristic timescale, $T$, as a function of $(b/a)^m$ for various cap sizes. In the limit of $nb/a\ll 1$, the timescale for all models goes as $aT\sim (b/a)^m$.}
    \label{fig:1}
\end{figure}

\subsection*{Cap model dynamics}

First, we consider so-called cap models, in which catastrophe is caused by the hydrolysis of at least $m$ out of the $n$ monomers nearest the growing (see Fig. 1). We refer to these $n$ monomers the ``cap region". Once the polymer experiences a catastrophe, it irrevocably depolymerizes back to length zero. (We note that microtubules do not strictly display the latter behavior, as they occasionally arrest their depolymerization and regrow before reaching zero length. However, complete depolymerization has been observed in bacterial actin homologs. By neglecting ``recovery and regrowth", we are in effect studying a previously reported ``bounded growth" phase \cite{dogterom1993,ranjinth2010}.) Such cap models have been explored in several previous works, which have calculated some of the model dynamics when $n=m$ \cite{brun2009, flyvbjerg1996}.

To discover the catastrophe time distribution in such models, we begin by enumerating the allowed states of the cap of the polymer (Fig. 1A). An individual monomer is either hydrolyzed ($\kcirc$) or unhydrolyzed ($\wcirc$), so that, e.g., the state space for $n,m=2$ is $\{\wcirc\wcirc,\wcirc\kcirc,\kcirc\wcirc,\kcirc\kcirc\}$. Within the constant rate models we consider here, the state $\wcirc\wcirc$ transitions to the states $\kcirc\wcirc$ or $\wcirc\kcirc$ by hydrolyzing at rate $b$; the state $\kcirc\wcirc$ transitions to $\wcirc\wcirc$ by polymerizing at rate $a$. The flow between these states is thus governed by a transition matrix, $\Mmat$, so that the probability of being in each state, $\pvec = (p_{\wcirc\wcirc},p_{\kcirc\wcirc},p_{\wcirc\kcirc},p_{\kcirc\kcirc})^T$, obeys the dynamical equation

\begin{equation}
\label{eq:dpdtCapModel}
\frac{d\pvec}{dt} = \Mmat\pvec \implies \pvec(t) = e^{\Mmat t}\pvec_0
\end{equation}

\noindent with $\pvec_0$ the initial probability state vector. Henceforth, we will refer to the last entry in the state vector as the absorbing state (with $m$ out of $n$ monomers hydrolyzed) by index $N$.

To find the catastrophe time distribution, $p_c(t)$, we may find the transition rate into the absorbing $N$th state (in which $m$ out of $n$ of the cap monomers are hydrolyzed):

\begin{equation}
\label{eq:pcCapModel}
p_c(t) = (\Mmat e^{\Mmat t}\pvec_0)_N.
\end{equation}

By numerically solving eq. \eqref{eq:pcCapModel} we observe (see Fig. 1C) that for polymers that grow to lengths $l\gg 1$, $p_c(t)$ is well-described by an exponential distribution with timescale $T\gg 1/a, 1/b$ (plus some transient behavior at times $t\approx 1/b$ that depends on the initial condition). This is true except in the case of $n,m=1$, for which the distribution is exactly exponential with timescale $1/b$, corresponding to the waiting time of a single Poisson event -- namely, the hydrolysis of the monomer at the tip.

To understand the nature of this nearly exponential distribution, we characterize the eigenspectrum of the transition matrix, $\Mmat$. We observe (see Fig. 1B) a constellation of eigenvalues in the complex plane. The negative real components of all but one of these eigenvalues are far from zero; it is this solitary small negative real eigenvalue that sets the timescale of the exponential distribution that dominates the catastrophe time distribution.

By plotting the catastrophe timescale, $T$, as a function of $(b/a)^m$, we see (Fig. 1D) that

\begin{equation}
\label{eq:TCapModel}
b/ma\ll 1\implies T\approx \frac{A_\text{cap}}{a}\left(\frac{a}{b}\right)^m,
\end{equation}

\noindent where $A_\text{cap}$ is an $n,m$-dependent constant (see Table 1). This relationship has been previously noted by Brun \textit{et al.} for the case of $n=m$ \cite{brun2009}. Here, we provide some intuition for how it comes about in general. Consider a process by which a monomer is added to the filament in every time interval of length $1/a$. At each time step, we draw a random configuration of the $n$ monomers in the cap; the monomer $k$ away from the tip has thus been in the filament for a time $k/a$; correspondingly, the probability that it has been hydrolyzed since its addition is $p_h(k)=1-e^{-kb/a}$. If we focus on $k\leq n$ with $nb/a\ll 1$, then $p_h(k)\approx kb/a$. The probability of encountering $m$ hydrolyzed monomers in the cap region thus scales as $(b/a)^m$ and the mean catastrophe time scales as in eq. \eqref{eq:TCapModel}. This intuition is not strictly rigorous because it treats the polymer state at each time step as independent from previous time steps. However, the state of the cap is completely erased after $n$ polymerization events and thus the assumption of independence should be accurate when the mean catastrophe time, $T$, greatly exceeds $na$.

\begin{figure}[t!]
    \centering
    \includegraphics[width=8.6cm]{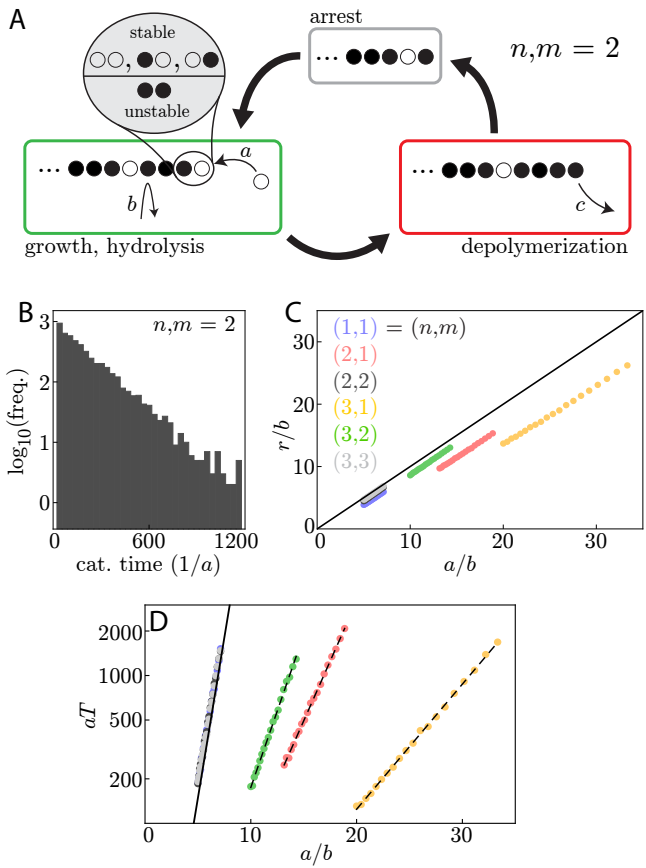}
    \caption{\textbf{Growth-and-shrinkage model dynamics.} \textbf{A:} Schematic of the growth-and-shrinkage model. Polymerization and hydrolysis proceed at constant rates of $a$ and $b$, respectively. If $m$ out of the first $n$ monomers become hydrolyzed, the filament becomes unstable and depolymerizes until it either finds a stable state (at which point it arrests depolymerization and regrows) or catastrophically depolymerizes. \textbf{B:} Numerically simulated histogram of 5000 catastrophe events for $n,m=2$ and $a/b=5$. As with cap models, growth-and-shrinkage models are governed by an exponential catastrophe time distribution. \textbf{C:} Numerically simulated elongation rate, $r/b$, as a function of the normalized polymerization rate, $a/b$. For the small $n,m$ growth-and-shrinkage models we tested, the effective elongation rate goes as $r/b = a/b-x$. The black line of $r/b=a/b$ is a guide to the eye. \textbf{D:} Numerically simulated catastrophe time, $bT$, as a function of the normalized polymerization rate $a/b$. Colors are the same as in panel \textbf{C}. Dashed black lines are exponential fits. The solid black line is the calculated mean catastrophe time using Markov chains, \eqref{eq:catTimeMarkov}. The mean catastrophe time grows exponentially in $a/b$ for every small $n,m$ model we tested.}
    \label{fig:2}
\end{figure}

{
\begin{table}
\centering
\begin{ruledtabular}
\begin{tabular}{c|c|c|c|c|c|c}
$(n,m)$ & $(1,1)$ & $(2,1)$ & $(2,2)$ & $(3,1)$ & $(3,2)$ & $(3,3)$ \\
\hline
\hline
$A_\text{cap}$ & 1 & 0.5 & 0.33 & 0.33 & 0.086 & 0.079 \\
\hline
$A_{gs}$ & 1.59 & 1.86 & 1.61 & 2.52 & 1.50 & 1.54 \\
\hline
$s$ & 0.96 & 0.37 & 0.96 & 0.19 & 0.48 & 0.96 \\
\hline
$x$ & 1.20 & 3.54 & 0.53 & 0.81 & 1.33 & 0.30 \\
\end{tabular}
\end{ruledtabular}

\caption{\textbf{Model parameters.} Extracted model parameters of $A_\text{cap}$ for the cap model and $A_{gs}$, $s$, and $x$ for the growth-and-shrinkage model.}
\label{table:dynamicssummary}
\end{table}
}

\subsection*{Growth-and-shrinkage model dynamics}

As appealing as cap models are for their conceptual simplicity, they fail to explain why a depolymerizing filament that encounters a stable cap region (say, $\wcirc\kcirc$, in the case of $n,m=2$) does not arrest its depolymerization and regrow. To that end, Antal \textit{et al.} have proposed a different class of growth-and-shrinkage models, in which a polymer tip may transition to a catastrophe state, depolymerize until it finds a stable cap region within the bulk of the polymer, arrest its depolymerization, and regrow (Fig. 2A) \cite{antal2007a,antal2007b}.

We assume that a filament is stable when a cap region of length $n$ has at most $m-1$ hydrolyzed monomers within it. In this state, the filament polymerizes at rate $a$. Each monomer in the filament hydrolyzes irreversibly at rate $b$. And when $m$ of the cap monomers hydrolyze, the polymer is destabilized. This instability causes depolymerization, which ceases only if the polymer depolymerizes into a stable cap region; if it fails to find such a stable region, it catastrophically shrinks to zero length. We assume, for mathematical convenience and in line with experiments in microtubules and actin homologs, that the depolymerization rate greatly exceeds the polymerization rate, and that the depolymerization process can thus be regarded as instantaneous. 

This cycle of constant growth and shrinkage changes the dynamics of the polymer as compared to the cap model. First of all, the apparent elongation rate of the polymer is no longer the polymerization rate, $a$, but some complex function $r(a,b,n,m)$. Secondly, the mean catastrophe time (the mean time from the beginning of growth to catastrophic depolymerization) is dramatically extended, owing to the fact that a depolymerizing filament can arrest the process by finding a stable cap region buried in the bulk.

To characterize the catastrophe time distribution, the mean catastrophe time, and the effective elongation rate, $r$, we performed simulations of the polymer growth and catastrophe process for various $n,m$ using the Gillespie Algorithm. Our findings are summarized in Figure 2. We show that the catastrophe time distribution of growth-and-shrinkage models also follows an exponential distribution, that the mean catastrophe time grows exponentially as $aT = A_{gs}\exp(sa/b)$ with $s$ and $A_{gs}$ both $n,m$-dependent constants (see Fig. 2 and eq. \eqref{eq:catTimeMarkov}), and that $r\approx a-bx$ with $x$ an $n,m$-dependent constant. See Table 1 for various values of $A_{gs}$, $s$, and $x$. All of these dependencies deviate strongly from the corresponding dynamics within the cap model.

We note that the mean catastrophe time of all models with $n=m$ is the same; this is because the catastrophe condition within these models is the same: every monomer must be hydrolyzed. We can calculate the relevant timescale by considering the probability of the polymer having $j$ unhydrolyzed monomers at a given time $t$, $p_j(t)$. States with $j$ unhydrolyzed monomers can make one of two transitions: (1) a monomer hydrolyzes and the polymer transitions into a state with $j-1$ unhydrolyzed monomers (this happens at rate $jb$) or (2) the polymer transitions to a state with $j+1$ unhydrolyzed monomers by polymerizing (at rate $a$). Thus, the dynamics are given by:

\begin{subequations}
\begin{equation}
dp_0/dt = bp_1
\end{equation}

\begin{equation}
dp_1/dt = 2bp_2-(a+b)p_1
\end{equation}

\begin{equation}
j\geq 2:~dp_j/dt = (j+1)bp_{j+1}+ap_{j-1}-(a+jb)p_j
\end{equation}
\end{subequations}

\noindent These dynamics again yield a transition matrix, $\mathbf{M}$, with a unique small negative real eigenvalue whose inverse is the exact mean catastrophe time. We can analytically calculate an approximate timescale by considering the mean time to the absorbing ($j=0$) state when starting from a singly unhydrolyzed ($j=1$) state. This timescale, $T_1$, can be calculated using properties of Markov chains \cite{taylor1998} and is given by

\begin{equation}
\label{eq:catTimeMarkov}
aT_1 = \sum_{k=1}^\infty \prod_{j=1}^k\frac{a}{jb} = \sum_{k=1}^\infty\frac{(a/b)^k}{k!} = e^{a/b}-1 \approx e^{a/b},
\end{equation}

\noindent which closely tracks the results from numerical simulation (see Fig. 2D) and the fit values of $s$ in Table 1.

Antal \textit{et al.} have previously calculated the mean catastrophe time and polymerization rate for the $n=m=1$ growth-and-shrinkage model by considering the probability that every monomer ever added to a given filament becomes hydrolyzed (a necessary condition for complete depolymerization when $n=m=1$) \cite{antal2007a}. Their relationship, $aT\approx \sqrt{\frac{a}{2\pi b}}e^{\pi^2a/6b}$, is roughly exponential in its functional form but treats the polymer state at each time step as independent from the previous time steps and thus overestimates the mean catastrophe time.

\subsection*{The universal distribution of constant-rate models}

The two models we have considered so far differ greatly in their mean catastrophe times and effective elongation rates. They rely on different underlying assumptions about the conditions that cause catastrophe. And yet, they display the same shape of the catastrophe time distribution. Why?

As we will now show, the shape of the catastrophe time distribution for models with constant polymerization and hydrolysis rates is universally exponential. Thus, the precise catastrophe condition is irrelevant; we ought always to expect an exponential distribution.

The reason for this universal behavior can be gleaned by considering a polymer that adds a monomer every $1/a$ time interval and has some catastrophe condition within a cap of size $n$. Suppose we happen upon the polymer and see it growing, but do not know its state. It has some probability $p_c$ of undergoing catastrophe prior to the next polymerization event. After $n$ polymerization events, the initial state of the cap is totally erased and its catastrophe probability prior to the next polymerization event is still $p_c$; the same thing is true after $n+1$ polymerization events. Thus, the catastrophe time distribution of the filament is simply the waiting time distribution of a single Poisson event, and is thus exponential for times $T> n/a$. (The condition $T>n/a$ is necessary to erase the polymer's ``memory" of the initial condition.) That this phenomenon is universal within a broad class of models justifies previous phenomenological theories, which assumed a constant catastrophe rate \cite{dogterom1993,bicout1997,hammele2003}.

In Appendix A, we show that this prediction is robust even to noise in the ratio of $a/b$, so that experiments that construct a catastrophe time distribution by binning events in slightly varying experimental conditions should still observe a roughly exponential distribution.

\begin{figure*}[t!]
    \centering
    \includegraphics[width=17.2cm]{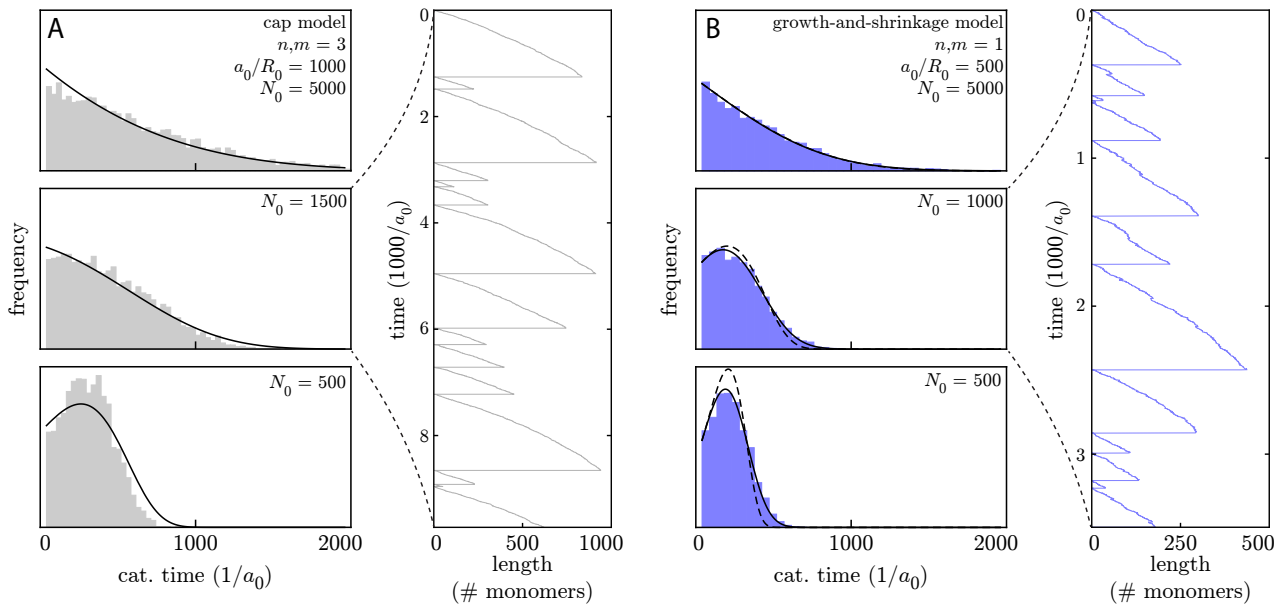}
    \caption{\textbf{Finite pool model dynamics.} \textbf{A:} Dynamics of finite pool cap models. \textit{Left:} Catastrophe time distributions with changing pool size, $N_0$. The histograms are numerical simulations of 5000 catastrophe events and the black line is eq. \eqref{eq:p(t)Cap}. The theory is expected to hold when $nb/a\ll 1$, which is roughly the case here ($n,m=3$, $b/a_0 \approx 0.043$). For pool sizes much larger than $a_0/R_0$, the catastrophe time distribution is approximately exponential. As the pool size shrinks, the catastrophe time distribution becomes distinctly non-exponential. \textit{Right:} Kymograph showing the filament length as a function of time with an initial pool size of $N_0=1500$. In these conditions, we can see a slowing down of the polymerization rate along with a change to the catastrophe time distribution. \textbf{B:} Dynamics of finite pool growth-and-shrinkage models. \textit{Right:} Catastrophe time distributions with changing pool size, $N_0$. The histograms are numerical simulations of 5000 catastrophe events; the solid black line is the nearly exact distribution, eq. \eqref{eq:p(t)GS}; and the dashed black line is the Gumbel-like approximate distribution, eq. \eqref{eq:p(t)GSLimit}. Here, we omit catastrophe events at times $t<b/a_0$ as these depend sensitively on the initial condition of the polymer. Again, we observe distinctly non-exponential catastrophe time distributions as the pool size, $N_0$, shrinks and becomes comparable with $a_0/R_0$. \textit{Left:} Kymograph showing the filament length as a function of time with an initial pool size of $N_0=1000$. The deviation from an exponential distribution is large, despite little apparent slowdown in the polymerization rate: in the conditions of the plot, a polymer of 250 monomers will polymerize at 75\% of the initial rate and undergo catastrophes at approximately 3.5 times the initial rate.}
    \label{fig:3}
\end{figure*}

\section*{Finite pool models}

So far we have treated the polymerization rate, $a$, as a constant. This seems a natural assumption given the roughly constant polymerization rate seen in experiments. However, as we have seen, the catastrophe timescale within certain models can be very sensitive to the ratio of $b/a$. Thus, it is natural to ask how the catastrophe time distribution is altered when monomers are depleted from a finite pool, thus lowering the polymerization rate. We should expect -- and will shortly demonstrate -- that alterations to the catastrophe distribution will be substantial when the initial pool size of $N$ monomers is roughly of order the mean catastrophe time of a polymer. Several previous works have also used numerical simulation and analytics to show qualitative changes to catastrophe dynamics within a different set of phenomenological models \cite{janulevicius2006, margolin2006}.

Here, we will treat polymerization as a first-order process in which the polymerization rate, $a$, is proportional to the local concentration of monomers in the pool. This is valid in the limit that monomers attach by simply ``running into" the tip of the growing polymer, which is thought to be the case for tubulin. This assumption appears to hold for microtubules, which show a linear dependence of elongation rate on tubulin concentration \cite{walker1988}. We will further assume that the finite pool is well-mixed, so that the polymerization rate is proportional to the total number of monomers in the pool, $N$. Protein diffusion constants are of order $D\sim 10^{-12}-10^{-11}$~m$^2$/s \cite{phillips2015}, so that for a typical polymerization rate of order $a\sim 1$/s, a monomer can diffuse a length $l\sim\sqrt{D/a}\approx 1~\mu$m of order the size of the cell; we thus expect the well-mixed assumption to hold. Finally, we will assume that the monomer pool is relatively large, so that each polymerization event only slightly changes the pool size.

We will first work out the general dynamics of such systems, then we will probe the specific dynamics for both the cap model and the growth-and-shrinkage model.

Once we specify a polymerization rate, a hydrolysis rate, and the polymer states that result in catastrophe, we can calculate the rate, $R(a,b)$, at which the polymer experiences catastrophes, which for the models we have studied here is given by the inverse of the mean catastrophe time: $R(a,b) = 1/T(a,b)$. The catastrophe rate acquires a time dependence through the ever-changing polymerization rate, $a(t)$. The probability that the filament survives without catastrophe until time $t$, $P_s(t)$, is thus governed by the following differential equation:

\begin{equation}
\label{eq:Ps}
\frac{dP_s}{dt} = -R(a(t),b)P_s.
\end{equation}

\noindent By formally integrating this equation and taking its time derivative, we can construct the catastrophe time distribution, $p_c(t) = -dP_s/dt$:

\begin{equation}
\label{eq:fpCatTimeDist}
p_c(t) = R(a(t),b)\exp\left[-\int_0^td\ttil~R(a(\ttil),b)\right].
\end{equation}

\noindent To find $p_c(t)$, we must therefore specify a catastrophe condition, which will determine the catastrophe rate, $R$. We must also determine the time dependence of the polymerization rate, $a$. Because $a$ is proportional to the monomer pool size, $N$, it suffices to specify an equation for $dN/dt$ -- how the pool size depletes over time. Both of these calculations are tractable within the models we have studied so far.

\subsection*{Finite-pool cap models}

Let's assume that the initial pool size is $N_0$ and that the initial polymerization rate is $a_0$. Thereafter, the polymerization rate $a$ will be given by $a(t)=a_0N(t)/N_0$. Within the cap model, the number of free monomers in solution decreases as

\begin{equation}
\label{eq:dNdtCap}
\frac{dN}{dt} = -a(t) = -a_0N(t)/N_0 \implies N(t) = N_0e^{-a_0t/N_0}.
\end{equation}

\noindent It follows that the polymerization rate, $a(t)$, is given by

\begin{equation}
\label{eq:a(t)Cap}
a(t) = a_0N(t)/N_0 = a_0e^{-a_0t/N_0}.
\end{equation}

\noindent We have previously seen that when $a/b\gg n$ for cap models with cap size, $n$, and critical damage number, $m$, the catastrophe rate is given by $R = 1/T = a(b/a)^m/A_\text{cap}$ with $A_\text{cap}$ a constant depending on $(n,m)$ (see Table 1). Thus, setting $R_0 = a_0(b/a_0)^m/A_\text{cap}$ and $R_1 = (m-1)a_0/N_0$, we have

\begin{equation}
\label{eq:R(t)Cap}
R(t) = R_0e^{R_1t}.
\end{equation}

\noindent We can therefore see that the catastrophe rate grows exponentially for cap models of $m>1$ within this limit. We can determine the catastrophe time distribution directly by plugging eq. \eqref{eq:R(t)Cap} into eq. \eqref{eq:fpCatTimeDist}:

\begin{equation}
\label{eq:p(t)Cap}
p(t) = R_0e^{R_0/R_1}e^{R_1t-\frac{R_0}{R_1}e^{R_1t}}.
\end{equation}

\noindent As we will see, this distribution (which takes on a Gumbel-like form) also governs the dynamics of some growth-and-shrinkage models. We emphasize that for cap models, it only applies when $a/b\ll n$; however, it captures some qualitative features of the actual catastrophe time distribution even when the agreement is not perfect (see Fig. 3A).

We have already noted that cap models have a catastrophe rate, $R$, that depends on the polymerization rate as $R\sim a^{1-m}$. The polymerization rate depends linearly on the pool size, $N$, so that $R\sim N^{1-m}$. For large $m$, $R$ can therefore depend very sensitively on $N$ -- much more sensitively than the linear dependence of $a\sim N$ -- and we can have large changes to the catastrophe time distribution with relatively small changes to the polymerization rate (see Fig. 3). 

\subsection*{Finite-pool growth-and-shrinkage models}

The processes underlying growth-and-shrinkage models deplete monomers from the pool at the effective elongation rate, $r$. As noted previously, this rate can differ from the polymerization rate, $a$, and is well-approximated by $r = a-bx$ with $x$ set by the cap parameters, $n$ and $m$ (see Table 1 and Fig. 2). Thus, we have that

\begin{multline}
\label{eq:N(t)GS}
\frac{dN}{dt} = -r(t) = -a(t)+bx \\
\implies N(t) = \frac{N_0}{a_0}\left[bx+(a_0-bx)e^{-a_0t/N_0}\right]
\end{multline}

\noindent while

\begin{equation}
\label{eq:a(t)GS}
a(t) = a_0N(t)/N_0 = bx+(a_0-bx)e^{-a_0t/N_0}.
\end{equation}

From our analysis of growth-and-shrinkage models we know that $A_{gs}R = ae^{-sa/b}$ with $A_{gs}$ and $s$ both $n,m$-dependent constants (see Table 1). Because $R$ depends so sensitively on $a/b$, a small decrease in $a/b$ due to depletion of the finite pool will cause a large increase in $R$ and quick catastrophe. We may therefore derive a nearly exact catastrophe time distribution by writing $a/b\approx a_0/b+\delta$ with $\delta\ll a_0/b$, which allows us to conclude that

\begin{multline}
\label{eq:RApproxGS}
A_{gs}R = ae^{-sa/b} = be^{-sa/b+\log\frac{a}{b}} \approx \\ a_0e^{-\frac{a}{b}(s-b/a_0)-1}.
\end{multline}

\noindent Setting $\stil=s-b/a_0$, $\Atil = eA_{gs}b/a_0$, $R_i = be^{-\stil x}/\Atil$, $R_1=a_0/N_0$, $y=\frac{\stil}{b}(a_0-bx)$, and $R_0=R_ie^{-y}$ and following the mechanics of the previous section yields the catastrophe time distribution

\begin{multline}
\label{eq:p(t)GS}
p_c(t) \approx R_i~\exp\bigg\{-ye^{-R_1t}\\
-\frac{R_i}{R_1}\left[\text{Ei}(-y)-\text{Ei}(-ye^{-R_1t})\right]\bigg\},
\end{multline}

\noindent with $\text{Ei}(.)$ the exponential integral function. We interpret $R_0$ as the initial catastrophe rate of the growing filament and $R_1$ as a reparameterized initial pool size.

At first glance, eqs. \eqref{eq:p(t)Cap} and \eqref{eq:p(t)GS} appear unrelated. However, when the initial polymerization rate divided by the initial catastrophe rate is much less than the initial pool size, we have $R_0\gg R_1$. In this limit, catastrophes occur overwhelmingly at times $t\sim 1/R_0\ll 1/R_1$ and the catastrophe rate simplifies to $R(t)\approx R_0e^{yR_1t}$ so that the catastrophe time distribution is given by

\begin{multline}
\label{eq:p(t)GSLimit}
R_0\gg R_1:~p_c(t) \approx R_0e^{R_0/yR_1}e^{yR_1t-\frac{R_0}{yR_1}e^{yR_1t}},
\end{multline}

\noindent exactly the same form as eq. \eqref{eq:p(t)Cap}. Thus, within certain limits, the finite-pool distributions of these models obey a common Gumbel-like shape. We explore the dynamics of the finite-pool catastrophe time distribution in Figure 3B, where we benchmark the dynamics of eqs. \eqref{eq:p(t)GS} and \eqref{eq:p(t)GSLimit} against simulations. As with finite-pool cap models, we observe that the removal of monomers from a finite pool can dramatically affect the catastrophe time distribution even when the polymerization rate is relatively unchanged. Actually, the effect is even stronger than for cap models, owing to the fact that while the polymerization rate goes like the pool size, the catastrophe rate is exponential in the polymerization rate (and hence the pool size).

\section*{Discussion}

In this work, we have elucidated universal and common features within models of dynamic instability. Such universal features are important for studies of biophysical phenomenon, as they offer predictions that are largely independent of microscopic details.

Experiments probing the catastrophe time distributions of microtubules have discovered that dynamic instability on the minus end is governed by an exponential distribution while the plus end is distinctly non-exponential \cite{odde1995}. This discrepancy is interesting from a modeling perspective as it supplies an opportunity to investigate the interplay of structure and dynamic instability. {Odde \textit{et al.} show that the observed non-exponential distribution can be explained by interaction between the quasi-one-dimensional microtubule protofilaments, each of which obeys an exponential catastrophe time distribution. The latter assumption also underlies well-known phenomenological models of dynamic instability \cite{dogterom1993,bicout1997,hammele2003}. We have shown that such an assumption is valid within a broad class of models.}

Other experimental observations -- including the adjustment of catastrophe times post-dilution -- have also been understood as arising from interactions between the one-dimensional protofilaments that comprise the microtubule \cite{flyvbjerg1996, brun2009}. Moreover, these interactions can be complicated by mechanical and chemical couplings \cite{Michaels2020, Brouhard2015, fletcher2010, vanburen2005}. It remains to be seen which of the features we have unearthed survive in these more complex settings. It would also be interesting to investigate theoretically the nucleation properties of dynamically unstable polymers \cite{oosawa1961, jonasson2020}.

Our work here focuses on dynamic instability of isolated one-dimensional polymers. It would be interesting to study the catastrophe and finite-pool dynamics of many dynamically unstable polymers growing simultaneously. Such studies could be compared with related studies of polymers that are not dynamically unstable \cite{mohapatra2017,fai2019}.

The complex geometrical structure of the microtubule makes it a difficult object to model. Direct application of our models may be more plausible for quasi-one-dimensional actin homologs \cite{garner2004, orlova2007}. By cultivating a deep understanding -- bridging theory and experiment -- of these simple polymers, we may begin to understand the precise details of the universal phenomena we have detailed here.

\section*{Acknowledgements}

PBD acknowledges support through the Paul M. Young Fellowship of the Fannie and John Hertz Foundation. JZ acknowledges support through The NSF-Simons Center for Mathematical and Statistical Analysis of Biology at Harvard (award \#1764269) and the Harvard Quantitative Biology Initiative.

\bibliography{bib}

\section*{Appendix A: Sensitivity analysis of the exponential distribution}

\begin{figure}[h!]
    \centering
    \includegraphics[width=8.6cm]{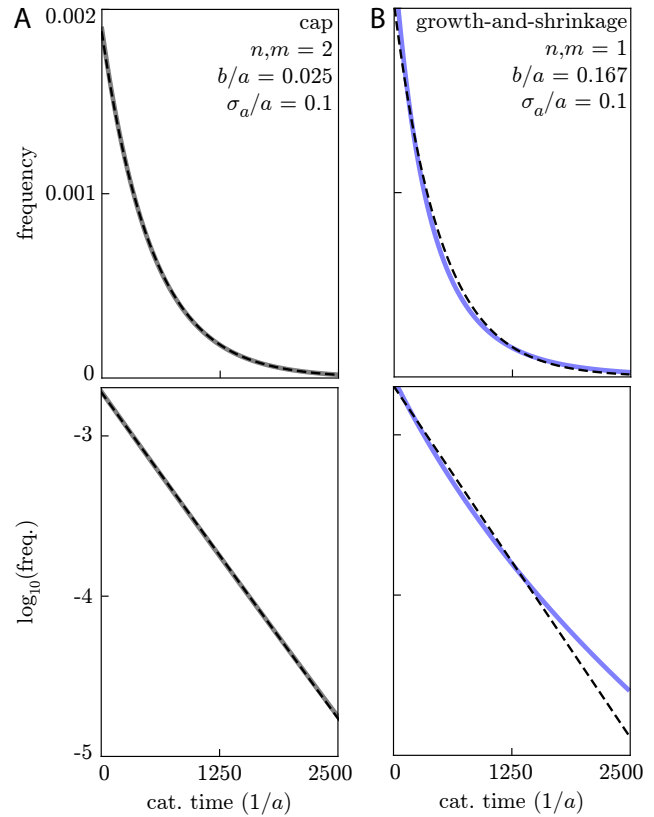}
    \caption{\textbf{Sensitivity analysis of the exponential distribution.} \textbf{A:} Catastrophe time distribution -- on linear (top) and semi-log (bottom) scales -- of a cap model with inter-experimental variation in the polymerization rate, $a$. We assume the polymerization rate is drawn from a gamma distribution with mean value, $a$, and standard deviation $\sigma_a = 0.1 a$. Even with added noise, the catastrophe time distribution (solid gray line) is almost exactly the same as the zero-variability distribution (dashed black line). \textbf{B:} Same as \textbf{A} but with the growth-and-shrinkage model. The distribution with noisy $a$ is given by the solid blue line while the dashed black line is the distribution with fixed $a$.}
    \label{fig:4}
\end{figure}

Here, we show that our main finding -- the ubiquity of the exponential catastrophe time distribution within constant-rate models -- is insensitive to noise. Imagine that you would like to observe the exponential catastrophe time distribution, but that the polymerization rate, $a$, changes between experiments. We will show that the exponential catastrophe time distribution is robust to such noise.

To do so, we assume that the polymerization rate in each experiment is drawn at random from some distribution, $q(a')$ with mean $a$ and standard deviation $\sigma_a$. With fixed $b$, the mean catastrophe time becomes a function of the particular value of $a'$ drawn from the distribution. Each $a'$ will correspond to an exponential distribution with mean catastrophe time $T(a',b)$, so that the empirical catastrophe time distribution will be given by

\begin{equation}
\label{eq:p(t)Noise}
p_c(t) = \int_0^\infty da'~q(a')\frac{e^{-t/T(a',b)}}{T(a',b)}.
\end{equation}

To check whether $p_c(t)$ is indeed well approximated by the noise-free distribution $e^{-t/T(a,b)}/T(a,b)$, we must make some assumptions about $q(a')$. For mathematical convenience, we pick $q(a')$ as a gamma distribution with mean $a=k\theta$ and variance $\sigma_a^2 = k\theta^2$. In Figure 3B, we show that under these assumptions, the noisy catastrophe time distribution differs only very slightly from the noise-free distribution for noise at the level of $\sigma_a/a = 0.1$.

\end{document}